\documentclass[aps,reprint,superscriptaddress,nofootinbib]{revtex4-1}

\usepackage{amsmath,graphicx}

\makeatletter
\def\sh{\mathop{\operator@font sh}\nolimits}
\def\ch{\mathop{\operator@font ch}\nolimits}
\makeatother
\def\be{\begin{equation}}
\def\ee{\end{equation}}

\begin{document}

\title{EPR = ER and Scattering Amplitude as Entanglement Entropy Change}

\author{Shigenori Seki}
\email[]{sigenori@hanyang.ac.kr}
\affiliation{Research Institute for Natural Science, Hanyang University, Seoul 133-791, Republic of Korea}

\author{Sang-Jin Sin}
\email[]{sjsin@hanyang.ac.kr}
\affiliation{Department of Physics, Hanyang University, Seoul 133-791, Republic of Korea}

%\date{\today}
\date{3 April 2014, Revised: 12 June 2014}

\begin{abstract}
We study the causal structure of  the minimal surface of the four-gluon scattering, 
and find a world-sheet wormhole parametrized by Mandelstam variables, thereby  
demonstrate the EPR = ER relation  for gluon scattering. 
We also propose that scattering amplitude is  the change of the 
entanglement entropy by generalizing  the holographic entanglement entropy 
of Ryu-Takayanagi to the case where two regions are divided 
in space-time. 
\end{abstract}

\pacs{}
%\keywords{}

\maketitle


\section{Introduction}

Quantum entanglement is one of the most subtle and intriguing property of the nature in the entire physics history.
When two pair-created particles fly away from each other, 
their states are entangled even after their separation is beyond causal contact. 
That is the Einstein-Podolsky-Rosen (EPR) pair \cite{EPR}. 
On the other hand, Einstein-Rosen bridge \cite{ER} connects far separated regions by short wormholes. 
Both of them shares the common nature such that 
causally disconnected objects or region are tied 
although no information can be transmitted  through them.
Recently Maldacena and Susskind \cite{MS} 
conjectured that any EPR pair  might be connected 
through a wormhole of some kind. It was dubbed as `EPR = ER'. 
If true, it would be a fascinating connection between quantum mechanics and 
space-time geometry giving an enlightenment on this long standing mystery 
of modern physics. 

Soon after this suggestion, Jensen and Karch \cite{JK} and Sonner
\cite{So} discussed the entanglement of a pair of accelerating quark and antiquark 
in the context of the AdS/CFT correspondence, 
using the corresponding minimal surface obtained by Ref.~\cite{Xi}. 
It allows one to consider only a classical world-sheet configuration 
where causal structure makes sense. 
It was shown that the trajectories of quark and antiquark are connected 
by a line that has to pass through the world-sheet wormhole zone, 
thereby supporting the EPR = ER with the space-time wormhole replaced by the world-sheet one. 
Ref.~\cite{CGP} suggested that the gluonic radiation between the quark and antiquark 
induces their entanglement. 
It is very interesting to see what happens to other exactly known world-sheet configuration \cite{JO,GP,Ni,HM}. 

In this paper we shall consider the four-gluon scattering, 
whose minimal surface was well studied by Alday and Maldacena \cite{AM}. 
We shall study its causal structure in its T-dual space-time picture and conclude 
that EPR = ER is also supported in this case.

Another related  question is how to quantify the  degree of entanglement. 
Notice that without interaction, 
unentangled state can not be entangled and vice versa.  
For example, in a scattering process of two particles starting with unentangled initial state, 
the final state is entangled if and only if there is an interaction, 
because the time evolution operator $U=\exp[-it (H_1+H_2+H_{\rm int})]$ factorizes iff $H_{\rm int}=0$.
The entanglement entropy (EE) of the final state  is the {\it change} of EE, 
$\Delta S_E$,  created by the interaction during the scattering process. 
So {\it the change of EE must be related to the interaction},  
hence we expect that the EE change is related to the scattering amplitude itself. 
In the AdS/CFT correspondence, the scattering amplitude can be related to 
the area of the minimal surface of the Wilson loop of trajectories of scattering particles \cite{RSZ,JP}, 
one way is to extend the EE derived 
from the minimal surface by Ryu and Takayanagi \cite{RT}. 
The relations between the EE 
and Wilson loop have been pointed out \cite{KP,LW,LM} for simple shape of the Wilson loop. 
We assume that the relation hold to more general cases. 
The gluon scattering amplitude was given from a polygonal Wilson loop in Ref.~\cite{AM}. 
Using all such data, we shall write down how these are connected.

\section{Minimal surface for gluon scattering}

Alday and Maldacena have considered the $AdS_5$ of momentum space, 
of which metric is denoted by
\begin{equation}
ds^2 = {R^2 \over r^2} \left(\eta_{\mu\nu}dy^\mu dy^\nu + dr^2 \right) \,, \quad 
\eta_{\mu\nu} = {\rm diag}(-1,1,1,1) \,, \label{adsmom}
\end{equation}
and have found the minimal surface solution corresponding to the gluon scattering \cite{AM}, 
\begin{eqnarray}
&&r = {\alpha \over \ch u_1 \ch u_2 + \beta \sh u_1 \sh u_2} \,, \nonumber \\
&&y_0 = r \sqrt{1+\beta^2} \sh u_1 \sh u_2 \,, \quad
y_3 = 0 \,, \nonumber \\
&&y_1 = r \sh u_1 \ch u_2 \,, \quad
y_2 = r \ch u_1 \sh u_2 \,,  \label{AMsolmom} 
\end{eqnarray}
where $\sh \equiv \sinh$ and $\ch \equiv \cosh$. 
$u_1$ and $u_2$ are the world-sheet coordinates. 
The boundary of this surface is a closed sequence of four light-like segments 
due to momentum conservation of gluons. 
$\alpha$ and $\beta$ are associated with Mandelstam variables \footnote{the Mandelstam variables are defined by $-s = (k_1 + k_2)^2 = 2k_{1\mu} k_2{}^\mu$, 
$-t = (k_1 + k_4)^2 = 2k_{1\mu} k_4{}^\mu$ and 
$-u = (k_1 + k_3)^2 = 2k_{1\mu} k_3{}^\mu = s+t$.} as 
\begin{equation}
-s\,(2\pi)^2 = {8\alpha^2 \over (1-\beta)^2} \,, \quad 
-t\,(2\pi)^2 = {8\alpha^2 \over (1+\beta)^2} \,, \label{sttoab}
\end{equation}
($0 \leq \beta \leq 1$).
In this paper we assume $s,t < 0$, that is to say, the $u$-channel. 
$\beta \to 1$ corresponds to the Regge limit, namely, 
$-s \to \infty$ with $-t$ fixed. 
Note that changing the sign of $\beta$ ({\it i.e.}, $-1 \leq \beta \leq 0$) 
is equivalent to exchanging $s$ and $t$.

We calculate the world-sheet induced metric on the surface (\ref{AMsolmom}), 
\begin{equation}
ds_{\rm ws}^2 = R^2 \left( du_1^2 + du_2^2 \right) \,, \label{momwsmet}
\end{equation}
and this induced metric is flat and Euclidean.

\begin{figure}
 \includegraphics[scale=0.9]{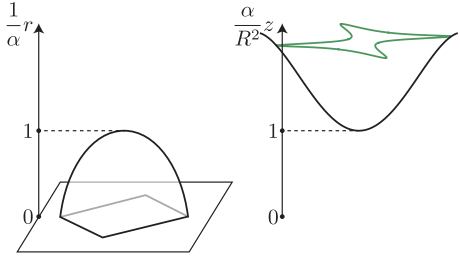}%
 \caption{\label{figtdual} The minimal surfaces in momentum space (left) and in position space (right).}
 \end{figure}

In order to obtain the surface for gluon scattering 
in the position space $(x^\mu, z)$, 
we use the ``T-dual'' transformation \cite{KT} (Fig.~\ref{figtdual}), 
\begin{equation}
\partial_m y^\mu = {R^2 \over z^2} \epsilon_{mn}\partial_n x^\mu \,, \quad 
z = {R^2 \over r} \,, \label{tdualtrf}
\end{equation}
so that the metric (\ref{adsmom}) is interpreted 
as an anti-de Sitter space again, 
$ds^2 = (R^2/z^2) (\eta_{\mu\nu}dx^\mu dx^\nu + dz^2 )$.
The transformation leads the solution (\ref{AMsolmom}) to 
\begin{eqnarray}
z &=& {R^2 \over 2\alpha}\left[ (1+\beta)\ch u_+ +(1-\beta)\ch u_-
\right] \,, \nonumber \\
x_0 &=& -{R^2 \over 2\alpha}\sqrt{1+\beta^2}\sh u_+ \sh u_- \,, \quad 
x_3 = 0 \,, \nonumber\\
x_+ &=& -{R^2 \over 2\sqrt{2}\alpha} \left[ (1+\beta)u_- +(1-\beta) \ch u_+
\sh u_- \right] \,, \nonumber\\
x_- &=& {R^2 \over 2\sqrt{2}\alpha} \left[ (1-\beta)u_+ +(1+\beta) \sh u_+
\ch u_- \right] \,, \label{solLC}
\end{eqnarray}
where we employed the space-time coordinates $x_\pm \equiv (x_1 \pm x_2)/\sqrt{2}$
and the world-sheet coordinates 
$u_\pm \equiv u_1 \pm u_2$ for convenience of calculation \cite{GPS}. 
Note that $x_\pm$ and $u_\pm$ are not light-cone coordinates and 
that $dx_1^2 + dx_2^2$ is equal to $dx_+^2 +dx_-^2$.

\section{Causal structure on world-sheet and entanglement}

The induced metric on the world-sheet (\ref{solLC}) in the position space is written down as 
\begin{equation}
ds_{\rm ws}^2 = R^2 \left( 
	g_{++}du_+^2 +2g_{+-}du_+ du_- +g_{--}du_-^2 
	\right) \,, \label{wsmetpm}
\end{equation}
with 
\begin{eqnarray}
g_{\pm\pm} &&= {2 \over \left[(1 +\beta)\ch u_+ +(1-\beta)\ch u_-\right]^2}\bigl[ (1\pm\beta)^2\sh^2 u_\pm \nonumber \\
&&+(1+\beta^2) -4^{-1}( (1\pm\beta)\ch u_\pm -(1\mp\beta)\ch u_\mp )^2 \bigr] \,, \nonumber \\
g_{+-} &&= {2(1-\beta^2)\sh u_+ \sh u_- \over \left[(1+\beta)\ch u_+ +(1-\beta)\ch u_-\right]^2} \,. \label{wsmetpmcomp} 
\end{eqnarray}
On this world-sheet there are two kinds of horizons: 
one is given by $g_{--} = 0$, {\it i.e.},
\begin{eqnarray}
 &&(1-\beta)\ch u_- +2\sqrt{(1-\beta)^2\sh^2 u_- +1 +\beta^2} \nonumber \\
&=& (1+\beta)\ch u_+ \,, \label{horminus}
\end{eqnarray}
and the other is given by $g_{++} = 0$, {\it i.e.},
\begin{eqnarray}
 && (1+\beta)\ch u_+ +2\sqrt{(1+\beta)^2\sh^2 u_+ +1 +\beta^2} \nonumber \\
&=& (1-\beta)\ch u_- \,. \label{horplus}
\end{eqnarray}
Note that the causal structure is induced in the world-sheet in position space 
by the ``T-dual'' transformation (\ref{tdualtrf}), although the world-sheet 
in momentum space (\ref{momwsmet}) is Euclidean. 
We introduce the rescaled coordinates, 
$X_\mu \equiv (\alpha/R^2) x_\mu$ ($\mu = 0,+,-,3$),  
and $Z \equiv (\alpha/R^2) z$.
Furthermore, in order to explicitly visualize the structure around infinity of $X_\pm$, 
we also use the coordinates, ${\hat X}_\pm \equiv (2 / \pi)\arctan X_\pm \in [-1,1]$. 
 \begin{figure}
 \includegraphics[scale=0.9]{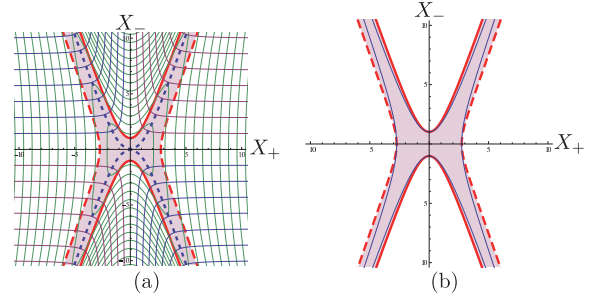}%
 \caption{\label{figsingularity} (a) The causal structure on the minimal surface in position space ($\beta = 1/2$). (b) The blue lines are the singularity.} 
 \end{figure}
 \begin{figure}
 \includegraphics[scale=0.9]{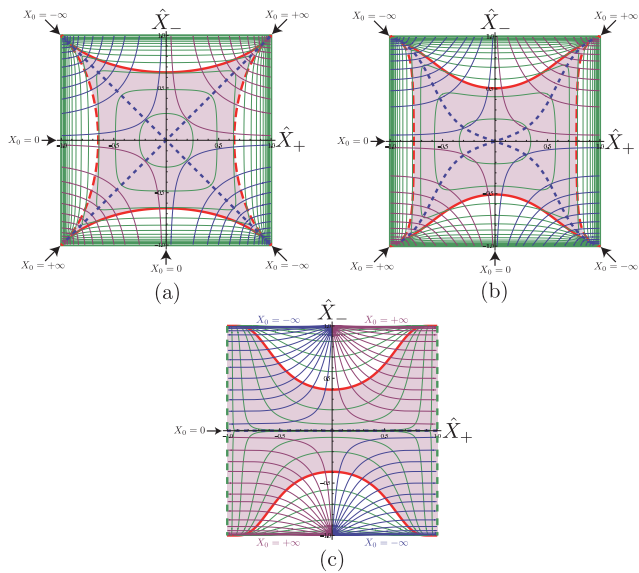}%
 \caption{\label{figcausaldiag}The causal structure on world-sheet. (a) $\beta = 0$, (b) $\beta = 1/2$, (c) $\beta = 1$.}
 \end{figure}
We depict the projection of minimal surfaces (\ref{solLC}) onto 
the $(X_+,X_-)$-plane in Fig.~\ref{figsingularity}(a) 
and the $({\hat X}_+,{\hat X}_-)$-plane in Fig.~\ref{figcausaldiag}.

Firstly we consider the case $0 \leq \beta < 1$.
Especially $\beta = 0$ implies that the scattering is symmetric 
with respect to $s$ and $t$ (see Eqs.~(\ref{sttoab})). 
The causal structure on world-sheet is drawn in Fig.~\ref{figcausaldiag}(a,b). 
The red solid lines are the horizons by $g_{--}=0$, {\it i.e.}, Eq.~(\ref{horminus}), and 
the red dashed lines are the horizons by $g_{++}=0$, {\it i.e.}, Eq.~(\ref{horplus}). 
In the red shaded regions, both of $g_{++}$ and $g_{--}$ are positive.
In every figure, $g_{++} >0$ and $g_{--} <0$ in the upper and lower white regions, 
while $g_{++} <0$ and $g_{--} >0$ in the left and right white regions. 
Therefore these white regions are Lorentzian, and are separated 
by the (red) Euclidean region, that is, a wormhole.

Note that $g_{--}$ is negative in the upper and lower Lorentzian regions, 
while $g_{++}$ is negative in the left and right Lorentzian regions, 
and that $g_{++}$ is equal to $g_{--}$ on the blue dotted lines given by 
$(1+\beta)\sh u_+ = \pm (1-\beta)\sh u_-$. 
It means that we can define world-sheet time as an appropriate coordinate depending on the region. 
Since the vertex operators can be inserted anywhere on the boundary of disk, 
this is completely natural. 
Consider a static gauge, $(\tau, \sigma) = (X_0, Z)$.
The time $\tau\, (=X_0)$ begins at the upper-left and lower-right corners and ends up 
at the upper-right and lower-left ones.
The thin blue (red) lines are negative (positive) constant $\tau$ lines. 
On the axes, $X_\pm =0$, $\tau$ is equal to zero. 
The thin green lines are constant $Z$ lines. 
The surface (\ref{solLC}) implies $Z \geq 1$. 
$Z$ has a minimum, $Z=1$, at the origin in Fig.~\ref{figtdual}(b) and Fig.~\ref{figcausaldiag}(a,b). 
$Z$ becomes infinity on the square bounding boxes, 
which are the AdS boundary \footnote{The minimal surface (\ref{solLC}) at 
$Z=\infty$ is laid on the AdS boundary, because simultaneously 
$X_\pm$ also goes to infinity (see Appendix A in Ref.~\cite{AM}).}, 
in Fig.~\ref{figcausaldiag}(a,b). 
Therefore we can recognize the thin blue and red lines as the 
time evolution of open strings whose endpoints 
are located on the AdS boundary.

The horizons (\ref{horminus}) and (\ref{horplus}) are at least the stationary limit curves 
but might be different from a horizon of usual black hole. 
So let us check whether there is a singularity. 
The Kretschmann scalar on the world-sheet (\ref{wsmetpm}), $R_{ijkl}R^{ijkl}$ ($i,j,k,l = \pm$), 
diverges on 
$(1-\beta)\ch u_- -(1+\beta)\ch u_+ = \pm 2\sqrt{1+\beta^2}$, 
in other words, these curves are singularity. 
From Fig.~\ref{figsingularity}(b), we can see that the singularity is in the interior of horizons,  
hence the horizons themselves are not singularity.

Next we focus on the case $\beta =1$. It is so-called the Regge limit, 
namely, $-s \to \infty$ with $-t$ fixed. 
The world-sheet metric (\ref{wsmetpm}) is reduced to 
\begin{equation}
R^{-2} ds_{\rm ws}^2 = \left({3 \over 2} -{1 \over \ch^2 u_+}\right)du_+^2 
	-\left({1 \over 2} -{1 \over \ch^2 u_+}\right) du_-^2 \,. \label{wsmetreg}
\end{equation}
While $g_{++}$ is positive definite, $g_{--}$ is negative when 
$\ch u_+ > \sqrt{2}$, {\it i.e.}, 
$|u_+| > \log (\sqrt{2} +1)$. 
Therefore the world-sheet horizons appear at 
$u_+ = \pm \log (\sqrt{2} +1)$. 
The causal structure on world-sheet is depicted in Fig.~\ref{figcausaldiag}(c), in which 
the red thick lines are the horizons given by $g_{--} = 0$. 
In this case different from those in $0 \leq \beta < 1$, 
two Lorentzian regions, where $g_{++} >0$ and $g_{--}<0$, exist, and 
are separated by a Euclidean wormhole (red shaded).

Since a gluon is described by an open string itself, 
we can see two kinds of entanglement: 
one is the entanglement of string endpoints in a gluon, 
and the other is the entanglement of gluons. 
 \begin{figure}
 \includegraphics[scale=0.9]{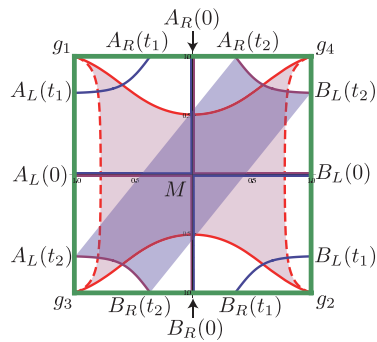}%
 \caption{\label{figentangle}The gluon scattering world-sheet projected onto $({\hat X}_+,{\hat X}_-)$. The boundary is denoted by the green box. The red region is a wormhole.}
 \end{figure}
In Fig.~\ref{figentangle}, $A_{L,R}$ and $B_{L,R}$ denote the endpoints of open 
strings describing gluons on the boundary. 
Since the upper-left and lower-right corners are at $X_0 = -\infty$ 
and the lower-left and upper-right corners are at $X_0 = \infty$, 
in the static gauge we can regard $g_1$ and $g_2$ as the incoming gluons 
and $g_3$ and $g_4$ as the outgoing gluons. 
The gluons, $g_1$ and $g_2$ at $X_0 = t_1\, (<0)$ and 
$g_3$ and $g_4$ at $X_0 = t_2\, (>0)$, can be described as the entangled states 
of open string endpoints, namely, 
\begin{eqnarray}
|g_1(t_1) \rangle\!\rangle &= \sum_{i,j} c^{(1)}_{ij}|A_{Li}(t_1)\rangle \otimes |A_{Rj}(t_1)\rangle \,,  \nonumber \\ 
|g_2(t_1) \rangle\!\rangle &= \sum_{i,j} c^{(2)}_{ij}|B_{Li}(t_1)\rangle \otimes |B_{Rj}(t_1)\rangle \,, \nonumber \\
|g_3(t_2) \rangle\!\rangle &= \sum_{i,j} c^{(3)}_{ij}|A_{Li}(t_2)\rangle \otimes |B_{Rj}(t_2)\rangle \,, \nonumber \\ 
|g_4(t_2) \rangle\!\rangle &= \sum_{i,j} c^{(4)}_{ij}|B_{Li}(t_2)\rangle \otimes |A_{Rj}(t_2)\rangle \,. \label{epentangle}
\end{eqnarray}
Each entanglement in the states (\ref{epentangle}) is interpreted to the fact 
that each open string crosses over the wormhole (see Fig.~\ref{figentangle}). 
Let us focus on the vicinities of the corners of (green) bounding box.  
The causal structure on the world-sheet which describes 
the entanglement of string endpoints in each gluon ({\it e.g.}~$A_R \to B_L$)
is similar to that of accelerating quark and antiquark in Ref.~\cite{JK} .

At $X_0 = 0$ the open strings, $g_1$ and $g_2$, join and split 
to $g_3$ and $g_4$, in other words, 
the color exchange of gluonic interaction happens at the mid-point $M$ of open strings (Fig.~\ref{figentangle}).  
Therefore $X_0 = 0$ is the moment that the entanglement 
between gluons is gained. 
Even if the initial state of gluons is not entangled, 
the final state of gluons is entangled due to the interaction. 
From a geometric viewpoint, any paths connecting the open string gluons 
({\it e.g.}~$A_R(t_2)B_L(t_2)$ and $A_L(t_2)B_R(t_2)$)
must cross the wormhole region (see the blue ribbon in Fig.~\ref{figentangle}).

\section{Entanglement entropy and scattering amplitude}
How can we quantify entanglement of two interacting particles? 
In Refs.~\cite{KP,LW,LM}, the EE is associated with a Wilson loop by 
$S_E = (1-c\lambda \partial_\lambda)\log \langle W \rangle$.\footnote{The undetermined constant $c$ 
depends on the shape of scattering Wilson loop and is not relevant to our purpose here. 
({\it cf.}~$c=4/3$  for a circular Wilson loop  Ref.~\cite{LM}.)} 
Note that the EE itself is associated with a quantum state
at a time while the Wilson loop $\langle W \rangle$ depends on the entire time dependent process. 
Therefore we should consider the left hand side of above mentioned equation 
as the {\it change} of the EE, $\Delta S_E$. 
So the gluon scattering amplitude \cite{AM} is related to the change 
of the EE in leading order of large $\lambda$ by 
\begin{eqnarray}
\Delta S_E 
&\sim& {(1-{1\over 2}c)\sqrt{\lambda} \over 8\pi}\left(\log{s \over t}\right)^2 \nonumber \\
&=& {(1-{1 \over 2}c)\sqrt{\lambda} \over 2\pi}\left(\log{1+\beta \over 1-\beta}\right)^2 \,, \label{EEscat}
\end{eqnarray}
where we neglected the IR divergent pieces.

We introduce another characteristic quantity concerning 
about the entanglement of gluons. 
Let us consider the proper lengths of lines, 
$A_R(0)B_R(0)$ and $A_L(0)B_L(0)$, at the contacting instance $X_0=0$; 
\begin{equation}
\ell_\pm(\beta) = R\int_{-u_{\pm\infty}}^{+u_{\pm\infty}} du_\pm \sqrt{g_{\pm\pm}}\big|_{u_\mp = 0} \,, \label{gluentlength}
\end{equation}
where we introduced the cutoff, $z_\infty\,(\gg 1)$, such that  
$(2\alpha / R^2) z_\infty = (1\pm\beta)\ch u_{\pm\infty}+1\mp\beta$.
 \begin{figure}
 \includegraphics[scale=0.9]{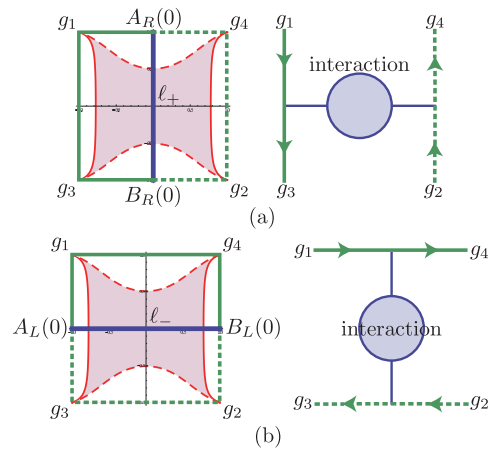}%
 \caption{\label{figgluent}The open string world-sheets and Feynman-like diagrams in the two channels of gluon interaction.}
 \end{figure}
$\ell_+$ and $\ell_-$ correspond to the two channels of gluon interaction. 
In one channel (Fig.~\ref{figgluent}(a)), the gluons $g_1$ and $g_2$ 
flow to $g_3$ and $g_4$ respectively. 
Then, the region $\Sigma$ on the boundary corresponding 
to the gluon $g_1 \to g_3$ is drawn by the thick green line segments 
and the region ${\overline \Sigma}$ corresponding 
to the gluon $g_2 \to g_4$ is drawn by the dotted green line segments. 
Since the blue line $A_R(0)B_R(0)$ in the bulk connects the boundary $\partial \Sigma$, $\ell_+$ is related with the entanglement of gluons 
in the sense of Ref.~\cite{RT}. 
In the same way, we can consider the other channel, {\it i.e.}, 
$g_1 \to g_4$ and $g_2 \to g_3$, 
in which $\ell_-$ characterizes a part of the entanglement of gluons (Fig.~\ref{figgluent}(b)).

Eq.~(\ref{gluentlength}) is computed as 
$\ell_\pm(\beta) = -\sqrt{6}R\log(1\pm \beta)$, 
where we subtracted the divergent piece, $\sqrt{6}R \log (2\alpha z_\infty/R^2)$, 
for $z_\infty \to \infty$. 
Then the EE change (\ref{EEscat}) is also described as 
\begin{equation}
\Delta S_E \sim {1-{1 \over 2}c \over 4\pi^{3/2}}\biggl({\ell_+ - \ell_- \over \ell_s} \biggr)^2 \,, \label{entfinite}
\end{equation}
where we used $R^2 = \sqrt{4\pi \lambda} \ell_s^2$.

We comment on the Regge limit, $\beta = \pm 1$. 
Since the finite part of $\Delta S_E$ becomes minimum at $\beta = 0$ and 
diverges at $\beta = \pm 1$, 
the Regge limit is the case with maximal $\Delta S_E$. 
Actually Fig.~\ref{figcausaldiag}(c) shows that, at $\beta =1$, 
one of the endpoints of $g_1$ ($g_2$) always coincides with that of $g_4$ ($g_3$), 
and $\ell_-(1)$ diverges.

Can we generalize above result to more general scattering particles? 
We believe this is the case. To show this we give a construction 
by which $\Delta S_E$ can be identified as the scattering amplitude. 
First we can extend the Ryu-Takayanagi formulation of EE 
by allowing the subspace A and B to be the space-time regions (rather than spatial regions) 
whose  minimal surface in AdS generates the change in the EE.  
In case of world-line of scattering quark-antiquark pair, 
it is nothing but the minimal surface calculating the Wilson lines. 
That is, for any   two scattering particles A and B, 
there is an infinite line $l(t)$ connecting them at each time $t$. 
 \begin{figure}
 \includegraphics[scale=0.9]{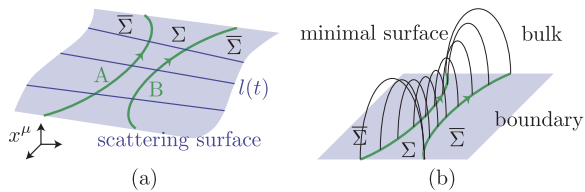}%
 \caption{\label{figscatsurface}(a) The scattering surface in the boundary theory. (b) The minimal surface of Wilson lines.}
 \end{figure}
As time evolves, $l(t)$ generates a two-dimensional surface 
in the entire space-time of boundary field theory, 
which we call a scattering surface (Fig.~\ref{figscatsurface}(a)).  
Then the world-lines of the two particles will divide the scattering surface 
into two, $\Sigma$ and ${\overline \Sigma}$.
Considering the constantly accelerating particles whose minimal surfaces is found in Ref.~\cite{Xi}, 
we can exemplify these idea. 
The trajectory of two particles forms a circle in Eudlideanized space-time. 
The minimal surface of circle is well studied, and its area is given 
by $-\sqrt{\lambda}/(2\pi)$ independent of the acceleration \cite{DGO}. 

This construction shows a way to identify the scattering amplitude 
as a change of EE.
Notice that the change of EE between initial and final states 
is a function of the whole scattering process. 
Therefore this change should be related to S-matrix.

The term, $\lambda \partial_\lambda \log \langle W \rangle$, in $\Delta S_E$ comes 
from the replica trick in the derivation of EE. 
On the other hand, in the language of scattering, that term corresponds to Bremsstralung 
of radiative correction. Actually in the case of accelerating quark-antiquark, 
$\lambda \partial_\lambda \log \langle W \rangle$ is 
proportional to the Bremsstralung function \cite{CHMS}. 
Therefore  the change of EE, $\Delta S_E$, is related 
with S-matrix, which gives a scattering amplitude in principle including a radiative correction.

\section{Conclusion}

We studied the causal structure on the open string world-sheet 
of gluon scattering minimal surface in position space. 
On this world-sheet there exists the wormhole 
which separates the Lorentzian regions including the boundary. 
Gluons are given by the open strings. 
We have shown that any paths connecting such two open string gluons 
at any time slices pass through the wormhole. 
Therefore a wormhole can always be associated 
with the entanglement of  interacting gluons.  
This result supports the EPR = ER conjecture.

Below, we discuss a few points which needs clarification: 
\begin{itemize}
\item One may ask why entanglement should be related to the interaction, 
because entanglement is property of the state not the hamiltonian.
Consider scattering of two particles 
which are initially (at $t=-\infty$) far separated and un-entangled.
We can construct the basis of in- and out- states by tensor product of free particle states. 
Let the initial state $|i\rangle$ to be a tensor product state $|i\rangle =|i_1\rangle \otimes |i_2\rangle $ 
They approach  each other and interact and then go force-free region after long time $ t=+\infty$. 
Such time evolution is given by the evolution operator 
$U=\exp[-iT (H_1+H_2+H_{\rm int})]$ or S-matrix: 
\begin{equation}
|\psi \rangle = \lim_{T\to \infty} U|i\rangle 
	= \sum _f |f\rangle\langle f|S|i \rangle 
	=\sum_f |f\rangle S_{fi}
\end{equation}
which is entangled in general unless interaction $H_{\rm int}=0$ so that $U$ is factorized.
So the final state of two free particles are entangled 
and its EE can be identified as the `change' of EE of the two particle system. 
Our question is that how to relate the latter to the S-matrix itself, 
which seems to be non-trivial task in field theory setting. 
%%%%%
\item If the final state involves sum over all possible quantum states, 
why one can consider only one world-sheet? 
In classical mechanics, final configuration is completely determined if initial one is given. 
Now in the AdS/CFT, due to the large $N$ nature, classical discussion can be made. 
That is, when we consider a minimal surface 
whose boundary is the trajectories of two scattering particles, 
we implicitly assumed that such classical picture is valid 
in describing the gluon-gluon or heavy quark-antiquark scatterings. 
Therefore we do not sum over trajectories and hence not sum over the world-sheets. 
This is the reason why we can consider the causal structure of a single world sheet 
of the gluon scattering instead of summing over such world-sheets. 
The same philosophy was implicitly assumed in the discussions of causal structure of world-sheet 
in recent literature. 
With these understanding, we observed the EE change (\ref{EEscat}), 
following the holographic calculations by Ref.~\cite{LM}.
This EE change becomes minimum at $\beta = 0$ and diverges at the Regge limit $\beta =\pm 1$. 
The relation shows that the change of EE is a function of dynamical process, which is natural. 
Here it was shown by holographic argument and mostly likely it is true only 
in the holographic context where semi-classical nature holds. 
It would be interesting  to  see how this relation 
in the general quantum field theory can be written. 
%%%%%
\item Another point that should be discussed further in the future is the conjecture we used: 
the EE of Wilson loop can be calculated by the minimal surface associated 
with the Wilson loop expectation value. Which was proven only simplest cases. 
Even providing more examples will be interesting. 
\end{itemize}

\begin{acknowledgments}
This work was supported by Mid-career Researcher Program 
through the National Research Foundation of Korea (NRF) grant No.~NRF-2013R1A2A2A05004846. 
SS was also supported in part by Basic Science Research Program through NRF grant No.~NRF-2013R1A1A2059434. 
SS is grateful to Institut des Hautes {\' E}tudes Scientifiques (IH{\' E}S) for their hospitality 
and to Thibault Damour and Robi Peschanski for helpful comments.   
SJS wants to thank Tadashi Takayanagi for illuminating discussions on EE at ICTP. 
\end{acknowledgments}


\begin{thebibliography}{25}%

\bibitem{EPR}
  A.~Einstein, B.~Podolsky and N.~Rosen,
  %``Can quantum mechanical description of physical reality be considered complete?,''
  Phys.\ Rev.\  {\bf 47} (1935) 777.
  %%CITATION = PHRVA,47,777;%%
  
\bibitem{ER}
  A.~Einstein and N.~Rosen,
  %``The Particle Problem in the General Theory of Relativity,''
  Phys.\ Rev.\  {\bf 48} (1935) 73.
  %%CITATION = PHRVA,48,73;%%

\bibitem{MS}
  J.~Maldacena and L.~Susskind,
  %``Cool horizons for entangled black holes,''
  Fortsch.\ Phys.\  {\bf 61} (2013) 781
  [arXiv:1306.0533 [hep-th]].
  %%CITATION = ARXIV:1306.0533;%%

\bibitem{JK}
  K.~Jensen and A.~Karch,
  %``The holographic dual of an EPR pair has a wormhole,''
  Phys.\ Rev.\ Lett.\  {\bf 111} (2013) 211602
  [arXiv:1307.1132 [hep-th]].
  %%CITATION = ARXIV:1307.1132;%%
  
\bibitem{So}
  J.~Sonner,
  %``Holographic Schwinger Effect and the Geometry of Entanglement,''
  Phys.\ Rev.\ Lett.\  {\bf 111} (2013) 211603
  [arXiv:1307.6850 [hep-th]].
  %%CITATION = ARXIV:1307.6850;%%
  
\bibitem{Xi}
  B.~-W.~Xiao,
  %``On the exact solution of the accelerating string in AdS(5) space,''
  Phys.\ Lett.\ B {\bf 665} (2008) 173
  [arXiv:0804.1343 [hep-th]].
  %%CITATION = ARXIV:0804.1343;%%
 
\bibitem{CGP}
  M.~Chernicoff, A.~G{\" u}ijosa and J.~F.~Pedraza,
  %``Holographic EPR Pairs, Wormholes and Radiation,''
  JHEP {\bf 1310} (2013) 211
  [arXiv:1308.3695 [hep-th]].
  %%CITATION = ARXIV:1308.3695;%%

\bibitem{JO}
  K.~Jensen and A.~O'Bannon,
  %``Holography, Entanglement Entropy, and Conformal Field Theories with Boundaries or Defects,''
  Phys.\ Rev.\ D {\bf 88} (2013) 106006
  [arXiv:1309.4523 [hep-th]].
  %%CITATION = ARXIV:1309.4523;%%

\bibitem{GP}
  H.~Gharibyan and R.~F.~Penna,
  %``Are entangled particles connected by wormholes? Support for the ER=EPR conjecture from entropy inequalities,''
  arXiv:1308.0289 [hep-th].
  %%CITATION = ARXIV:1308.0289;%%
    
\bibitem{Ni}
  H.~Nikolic,
  %``Can a wormhole be interpreted as an EPR pair?,''
  arXiv:1307.1604 [hep-th].
  %%CITATION = ARXIV:1307.1604;%%

\bibitem{HM}
  T.~Hartman and J.~Maldacena,
  %``Time Evolution of Entanglement Entropy from Black Hole Interiors,''
  JHEP {\bf 1305} (2013) 014
  [arXiv:1303.1080 [hep-th]].
  %%CITATION = ARXIV:1303.1080;%%

\bibitem{AM}
  L.~F.~Alday and J.~M.~Maldacena,
  %``Gluon scattering amplitudes at strong coupling,''
  JHEP {\bf 0706} (2007) 064
  [arXiv:0705.0303 [hep-th]].
  %%CITATION = ARXIV:0705.0303;%%

\bibitem{RSZ}
  M.~Rho, S.~-J.~Sin and I.~Zahed,
  %``Elastic parton-parton scattering from AdS / CFT,''
  Phys.\ Lett.\ B {\bf 466} (1999) 199
  [hep-th/9907126].
  %%CITATION = HEP-TH/9907126;%%

\bibitem{JP}
  R.~A.~Janik and R.~B.~Peschanski,
  %``Minimal surfaces and Reggeization in the AdS / CFT correspondence,''
  Nucl.\ Phys.\ B {\bf 586} (2000) 163
  [hep-th/0003059].
  %%CITATION = HEP-TH/0003059;%%

\bibitem{RT}
  S.~Ryu and T.~Takayanagi,
  %``Holographic derivation of entanglement entropy from AdS/CFT,''
  Phys.\ Rev.\ Lett.\  {\bf 96} (2006) 181602
  [hep-th/0603001].
  %%CITATION = HEP-TH/0603001;%%

\bibitem{KP}
  A.~Kitaev and J.~Preskill,
  %``Topological entanglement entropy,''
  Phys.\ Rev.\ Lett.\  {\bf 96} (2006) 110404
  [hep-th/0510092].
  %%CITATION = HEP-TH/0510092;%%

\bibitem{LW}
  M.~Levin and X.~-G.~Wen,
  %``Detecting Topological Order in a Ground State Wave Function,''
  Phys.\ Rev.\ Lett.\  {\bf 96} (2006) 110405.
  %%CITATION = PRLTA,96,110405;%%

\bibitem{LM}
  A.~Lewkowycz and J.~Maldacena,
  %``Exact results for the entanglement entropy and the energy radiated by a quark,''
  JHEP {\bf 1405} (2014) 025
  [arXiv:1312.5682 [hep-th]].
  %%CITATION = ARXIV:1312.5682;%%

\bibitem{KT}
  R.~Kallosh and A.~A.~Tseytlin,
  %``Simplifying superstring action on AdS(5) x S**5,''
  JHEP {\bf 9810} (1998) 016
  [hep-th/9808088].
  %%CITATION = HEP-TH/9808088;%%

\bibitem{GPS}
  M.~Giordano, R.~Peschanski and S.~Seki,
  %``Eikonal Approach to N=4 SYM Regge Amplitudes in the AdS/CFT Correspondence,''
  Acta Phys.\ Polon.\ B {\bf 43} (2012) 1289
  [arXiv:1110.3680 [hep-th]].
  %%CITATION = ARXIV:1110.3680;%%

\bibitem{DGO}
  N.~Drukker, D.~J.~Gross and H.~Ooguri,
  %``Wilson loops and minimal surfaces,''
  Phys.\ Rev.\ D {\bf 60} (1999) 125006
  [hep-th/9904191].
  %%CITATION = HEP-TH/9904191;%%
  
\bibitem{CHMS}
  D.~Correa, J.~Henn, J.~Maldacena and A.~Sever,
  %``An exact formula for the radiation of a moving quark in N=4 super Yang Mills,''
  JHEP {\bf 1206} (2012) 048
  [arXiv:1202.4455 [hep-th]].
  %%CITATION = ARXIV:1202.4455;%%

\end{thebibliography}
\end{document}